\documentclass[letterpaper, 10 pt, conference]{ieeeconf}  

\IEEEoverridecommandlockouts                              

\usepackage[ruled,vlined]{algorithm2e}
\usepackage[hyphens]{url}
\usepackage{hyperref}
\newenvironment{myitemize}{\begin{list}{$\bullet$}
{\setlength{\topsep}{1mm}
\setlength{\itemsep}{0.25mm}
\setlength{\parsep}{0.25mm}
\setlength{\itemindent}{0mm}
\setlength{\partopsep}{0mm}
\setlength{\labelwidth}{15mm}
\setlength{\leftmargin}{4mm}}}{\end{list}}

\usepackage{array}
\newcommand{\PreserveBackslash}[1]{\let\temp=\\#1\let\\=\temp}
\newcolumntype{C}[1]{>{\PreserveBackslash\centering}p{#1}}
\newcolumntype{R}[1]{>{\PreserveBackslash\raggedleft}p{#1}}
\newcolumntype{L}[1]{>{\PreserveBackslash\raggedright}p{#1}}

\usepackage{graphics} 
\usepackage{epsfig} 
\usepackage{mathptmx} 
\usepackage{times} 
\usepackage{amsmath} 
\usepackage{amssymb}  
\usepackage{multirow}
\usepackage{bm}
\usepackage{graphicx}
\usepackage{xcolor}
\usepackage{diagbox}

\title{\LARGE \bf Securing Connected Vehicle Applications with an Efficient \\ Dual Cyber-Physical Blockchain Framework}

\author{Xiangguo Liu$^{1}$, Baiting Luo$^{1}$, Ahmed Abdo$^{2}$, Nael Abu-Ghazaleh$^{2}$, Qi Zhu$^{1}$
\thanks{$^{1}$Xiangguo Liu, Baiting Luo and Qi Zhu are with the Department of Electrical and Computer Engineering, Northwestern University, Evanston, IL 60201, USA.
        {\tt\small xg.liu@u.northwestern.edu, baitingluo2019@u.northwestern.edu, qzhu@northwestern.edu.}}%
\thanks{$^{2}$Ahmed Abdo and Nael Abu-Ghazaleh are with the Department of Electrical and Computer Engineering, University of California, Riverside, CA 92521, USA.
        {\tt\small aabdo003@ucr.edu, nael@cs.ucr.edu.}}%
}

\begin{document}


\maketitle


\thispagestyle{empty}
\pagestyle{empty}

\begin{abstract}
While connected vehicle (CV) applications have the potential to revolutionize traditional transportation system, cyber and physical attacks on them could be devastating. In this work, we propose an efficient dual cyber-physical blockchain framework to build trust and secure communication for CV applications. Our approach incorporates blockchain technology and physical sensing capabilities of vehicles to quickly react to attacks in a large-scale vehicular network, with low resource overhead. We explore the application of our framework to three CV applications, i.e., highway merging, intelligent intersection management, and traffic network with route choices. Simulation results demonstrate the effectiveness of our blockchain-based framework in defending against spoofing attacks, bad mouthing attacks, and Sybil and voting attacks. We also provide analysis to demonstrate the timing efficiency of our framework and the low computation, communication, and storage overhead for its implementation\footnote{This work has been submitted to the IEEE for possible publication. Copyright may be transferred without notice, after which this version may no longer be accessible.}.
\end{abstract}

\section{Introduction}


Connected vehicle (CV) applications are expected to revolutionize traditional transportation system. Connected vehicles exchange messages with surrounding vehicles and infrastructure units for extending perception range, exchanging traffic status in downstream, and coordinating planning and control to improve safety, economy, and traffic flow. 
The U.S. Department of Transportation (USDOT) has identified a number of promising CV applications and started deploying them at three test sites in Florida, New York, and Wyoming~\cite{intelligent_transportation_systems}. Many other academic and industry testbeds and deployments are also under way~\cite{connected_vehicle_test_bed}. However, cyber and physical attacks targeting these systems can have severe safety implications, causing accidents or disrupting traffic flow. For example, an intelligent intersection management system can experience deadlock if messages have a long transmission delay or get lost under the denial of service attack~\cite{chen2018exposing,zheng2019design}. Merging in the highway will be more prone to accidents if vehicles get wrong position data of surrounding vehicles under spoofing and false message attacks~\cite{abdo2019application}. As the impact of cyber threats~\cite{parkinson2017cyber} on CV operations can be so destructive, it is essential to develop security solutions against them.   


Research in the literature has investigated possible methods to mitigate attacks on CV applications. Cryptographic techniques, message authentication, and digital signatures are widely used for computer networks, and can be leveraged for vehicular communication networks~\cite{hamoud2017security}. Graph-based methods~\cite{yu2006sybilguard,danezis2009sybilinfer,yu2008sybillimit} utilized in social networks may be used to distinguish attackers from honest vehicles by finding a cut between their representations in the graph.
Recent advances in hardware-assisted security are also leveraged to ensure the integrity of the sensor data in CVs~\cite{hu2020cvshield}.

A central issue in CV security is to build trust among vehicles. The authors in~\cite{gurung2013information} reviewed methods to evaluate message trustworthiness in the vehicular network, including entity-oriented, data-centric, and collaborative trust models. The entity-oriented trust models evaluate messages' trustworthiness based on the trustworthiness of their senders. A Certification Authority is often leveraged to record vehicles' behavior and provide estimations of trust; otherwise, a vehicle needs to collect information and make evaluation by itself. The data-centric trust models evaluate the content of messages to estimate the trust for them. In this category, Bayesian inference, and Dempster-Shafer theory are popular methods to estimate the plausibility and trustworthiness of messages, 
however they may result in false positives for detecting valid messages~\cite{raya2008data}. 
The collaborative trust models are based on integrating trust estimates from other peer vehicles, which could be time-consuming and attacked by malicious players.  
The Security Credential Management System (SCMS) is a proof-of-concept message security solution that is supported and developed by USDOT. Instead of evaluating message trustworthiness by each vehicle, message senders with credentials can be trusted in the system. However, since a certified vehicle can be attacked later, it is challenging for certificate authorities to track and update vehicles' status quickly. Thus, SCMS on its own does not prevent application level attacks~\cite{abdo2019application}.


Blockchain~\cite{kolb2020core} technology has natural strength in recording transactions/events and reaching a secure consensus among all users. It provides a promising direction for building trust in CV applications, as shown in~\cite{iqbal2019trust,singh2018branch,yang2018blockchain}. 
However, the required secure consensus operations in those earlier works introduce high overhead that makes it difficult for applying them in practical CV applications. 

\begin{figure}[t]
\centering\includegraphics[scale=0.3]{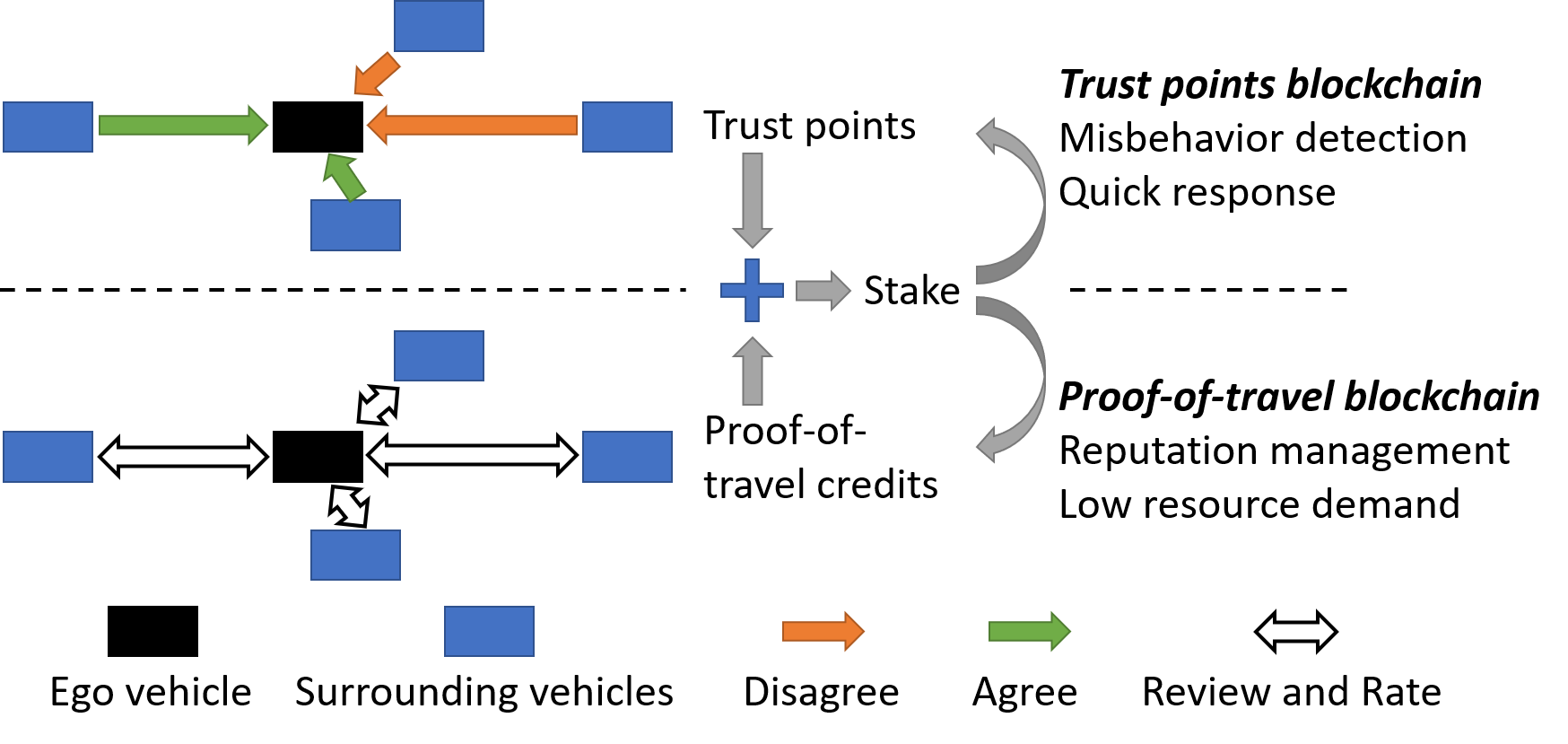}
\caption{Overview of our proposed dual cyber-physical blockchain framework. \textmd{The ego vehicle is traveling and sharing traffic information with surrounding vehicles. In the trust points blockchain, surrounding vehicles leverage their physical sensing capability to verify the messages sent from the ego vehicle. If a falsified message is detected and reported, the voting contract will instantly start collecting surrounding vehicles' opinions and adjust trust points of relevant vehicles. In the proof-of-travel blockchain, ego vehicle and surrounding vehicles record the number of received messages from different vehicles, which can reflect the travel activities and their contributions. To reduce resource overhead, the credits are updated in a longer period than that of trust points blockchain for reflecting long-term reputation. The stake of each vehicle is computed from both its trust points and proof-of-travel credits. Our dual blockchain design is secure as long as more than 2/3 of the stake is held by honest vehicles. 
}}
\label{fig:overview}
\end{figure}

In this work, we present an efficient dual cyber-physical block-chain framework to build trust and secure communication for CV applications. The framework enables us to efficiently track and update a trust estimate for each vehicle in a large-scale traffic network, with low resource overhead.  
The design of our framework is shown in Fig.~\ref{fig:overview}. It builds and updates two blockchains for recording vehicles' communication activities and establishing trust among vehicles. 
The first blockchain, named \emph{trust points blockchain}, is used to quickly identify and record malicious misbehavior. Intuitively, telling the truth and getting acknowledged by most neighbors will earn trust points, while telling a lie will lose trust points. The second blockchain, named \emph{proof-of-travel blockchain}, accumulates and records each vehicle's long-term contribution 
to the CV community. 
Intuitively, the more traffic information it shares with others, the more contribution to the CV community it gets and the higher proof-of-travel credits it can receive from others.  This blockchain serves as a low-cost conceptually-centralized tracker of vehicle trust, enabling two vehicles to establish trust when they do not have extended experience with each other, in a way that is secure and dependent on consensus.

Our proposed framework makes it difficult to launch attacks, and facilitates quick detection and reaction to misbehavior. Specifically, once a suspicious message is reported, surrounding vehicles can adapt to more cautious and conservative actions within just hundreds of milliseconds. Within one minute, surrounding vehicles can leverage their sensing capabilities to verify the message and reach consensus on it. Then, the trust points of the vehicle that sent the suspicious message will get updated in the trust points blockchain. On the other hand, as travel history is recorded in the proof-of-travel blockchain, vehicles with poor travel history cannot easily start certain attacks such as Sybil attack or flooding attack (or other attacks that require repeated transmission).   
Moreover, to improve the efficiency, we propose the concepts of permanent address and current active address, which can partition all vehicles into different regions with the sharding technique. Our framework design enables record transfer when vehicles move across different regions. Each region can have its blockchains updated and maintained with intra-region and inter-region communication to promote scalability. 

Our contribution in this paper can be summarized as follows:
\begin{myitemize}
\item We develop a novel dual blockchain framework that leverages cybersecurity techniques, physical sensing capabilities of vehicles, and their travel histories to build trust and secure communication in a large-scale vehicular network, in which every vehicle can get updates timely with low resource overhead.
\item We use a stake-based consensus mechanism across the faster trust points blockchain and the slower proof-of-travel blockchain, a sharding technique for partitioning vehicles into regions, and a dedicated summary step to reduce computation, communication and storage costs of our framework so that it is practical for CV applications. 
\item We demonstrate the effectiveness of our framework against  a few prevalent attacks, including message spoofing attack, bad mouthing attack, and Sybil attack. We illustrate the performance and timing efficiency of our defense system via simulations in the SUMO~\cite{lopez2018microscopic} simulator, and also
analyze its resource overhead and trade-offs. 
\end{myitemize}

The rest of the paper is organized as follows. Section~\ref{sec: background and related work} briefly introduces blockchain and Algorind, a recent blockchain design that we leverage in this paper, 
and previous work in applying blockchain to transportation systems. Section~\ref{sec: methodology} presents the design of our dual blockchain framework. Section~\ref{sec: defense} analyzes our framework's defense performance against message spoofing attack, bad mouthing attack, and Sybil and voting attack, and presents the simulation results in SUMO. Section~\ref{sec: resource demand} analyzes the computation, communication and storage overhead for each vehicle, and the trade-offs in adjusting the region size that one blockchain covers and the block generation period. Section~\ref{sec: conclusion} concludes the paper.

\section{Background and Related Work}\label{sec: background and related work}

\subsection{Threat Models to Connected Vehicles}\label{subsec:threat_model}


In this work, we focus on the attacks from vehicle-side devices, e.g., those from the On-Board Units (OBUs) for CV applications~\cite{chen2018exposing}. While the techniques can be extended to protect against attacks on infrastructure units, it is reasonable to expect that it is typically harder to attack the infrastructure. 
In particular, we consider malicious vehicles that can generate falsified messages and broadcast them 
to other vehicles. It is important to note that we do not assume that the attacker can spoof the sender identities in the messages. We assume that such an identity is verifiable and non-forgeable with digital signatures technique (e.g., SCMS). Moreover, we assume that the communication infrastructure is mostly resilient and most surrounding vehicles within a known communication range should receive the same message within a known time bound. We also assume that most honest vehicles have loosely synchronized clocks (e.g., using Network Time Protocol (NTP)) for the liveness of blockchain. Finally, we assume that most vehicles in the traffic network are honest, such that most of the stake (more than 2/3) can be held by honest vehicles to ensure the safety of blockchain. 

\begin{figure}[t]
\centering\includegraphics[scale=0.26]{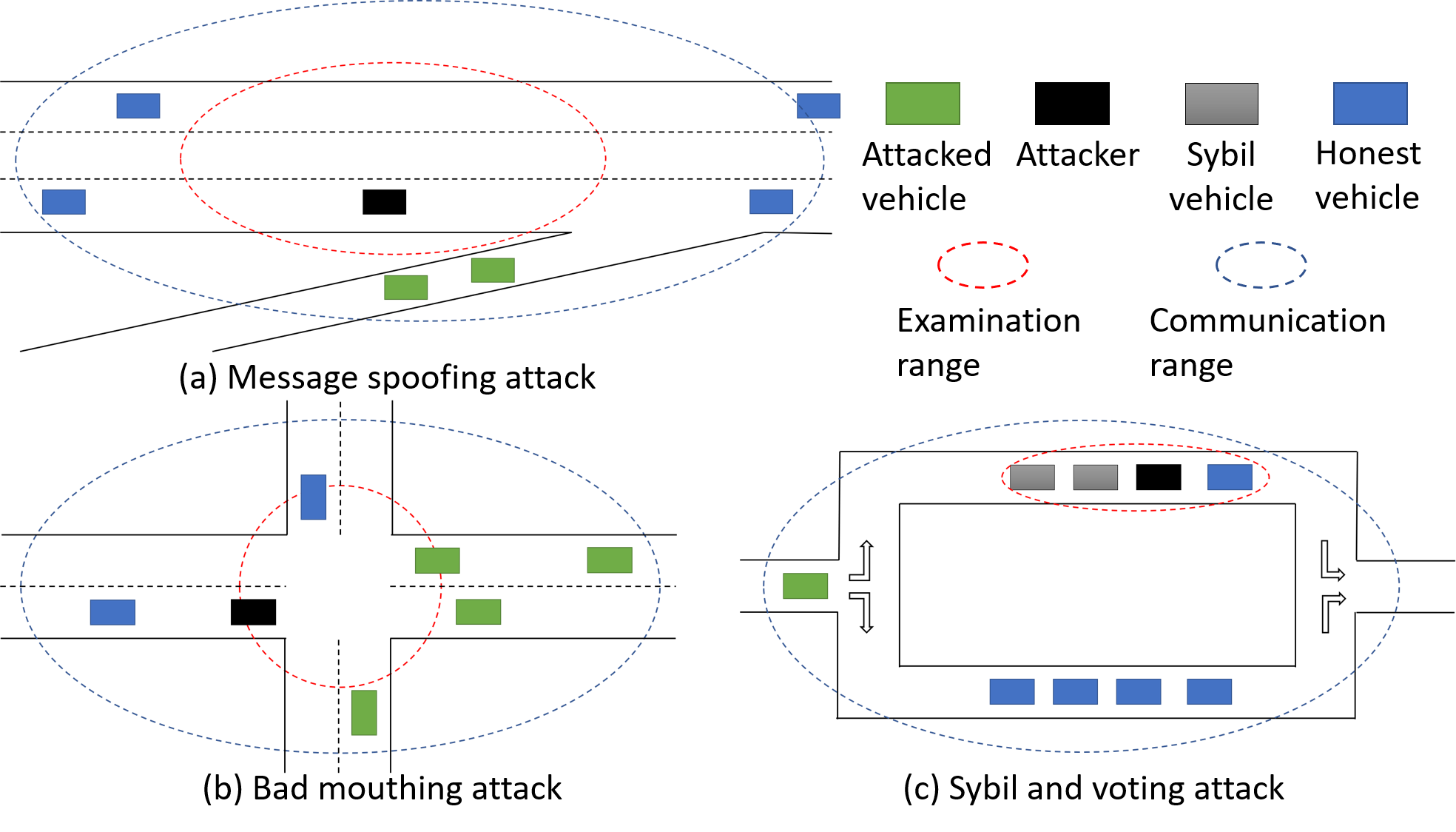}
\caption{Threat models considered in our work. \textmd{Subplot~(a) displays that an attacker sends falsified messages with wrong traffic status on the highway to mislead the green merging vehicles. Subplot~(b) displays that an attacker forges misbehavior of surrounding honest vehicles to hurt their reputation and traffic efficiency in the intelligent intersection. Subplot~(c) displays that an attacker forges two pseudonymous identities by launching a Sybil attack. Together they report falsified traffic accidents in one of the routes, and may dominate the voice in voting among the surrounding vehicles so that the falsified messages would not be detected. This attack may lead to larger traffic density and travel time because more honest vehicles will choose the other route that is without the fake accident.}}
\label{fig_attacks}
\end{figure}

Under these assumptions, malicious attackers can compromise a small fraction of vehicles to broadcast falsified position and velocity data, forge traffic accidents and destroy others' reputation. 
Specifically, we consider three threat models in this paper with popular and representative CV applications, as shown in Fig.~\ref{fig_attacks}. Among them, the message spoofing attack could target many other CV applications and be defended by trust frameworks such as ours. The bad mouthing attack and Sybil and voting attack in fact target the trust framework itself, and as shown later, our blockchain framework can effectively mitigate such attacks as well.    

\subsubsection{Message spoofing attack}
In Fig.~\ref{fig_attacks}(a), two green vehicles communicate with the vehicles on the highway to adjust their speed for merging onto the highway. The attacker (shown in black) can send out falsified position and velocity data of itself, which may induce the green victim vehicles to accelerate/decelerate. This may damage traffic efficiency and even put vehicles near the on-ramp at the risk of collisions. A single malicious vehicle may start this type of attack when no honest vehicle can judge the truth of the message or the stake of attackers is overwhelming in the local area.

\subsubsection{Bad mouthing attack}
In Fig.~\ref{fig_attacks}(b), all vehicles send their current and destination lanes as well as estimated arriving time to a centralized intersection manager, which then decides the passing order and timing of the vehicles, similarly as in~\cite{zheng2019design}. 
When a trust framework is deployed to secure the communication, an attacker may maliciously report misbehavior to degrade the trust for honest vehicles~\cite{sun2006trust}. Without an accurate assessment of the trust, the intersection manager will not be able to decide the passing order and timing for vehicles, and they will have to adapt the traditional traffic rules as if there were stop signs in every direction.   

\subsubsection{Sybil and voting attack}
In Fig.~\ref{fig_attacks}(c), there are two route choices in the region. 
An attacker can generate fake and pseudonymous vehicles by Sybil attack. Together they can broadcast a falsified traffic accident in one of the routes. In a trust framework based on consensus, while honest vehicles may disagree with the falsified accident, their voices could be dominated by the attacker and Sybil vehicles in the voting process for assessing the falsified accident, in which case, other vehicles coming to this region will believe the falsified accident and go through the other route, resulting in congestion and low traffic efficiency.  


\subsection{Blockchain and Algorand}


Blockchain is literally a chain of blocks (called a ledger), in which a block stores a set of transactions and the hash value of the previous block. Transactions are public to every user, and thus blocks can be verified by every user. Blockchain is updated based on consensus algorithms to guarantee that the state of the ledger is identical at all nodes. The most well-known blockchain, bitcoin, is based on proof of work. It assumes that a group of miners will build up new blocks by consuming CPU resources to find the right cryptographic nonce. Once a new block is proposed by a miner, other miners will verify it and start working on the next block. 
Because the block generation process is not deterministic, a hacker may successfully build a new block with a slight possibility. In this way, a block is considered to be confirmed if it is followed by several other blocks. 
For bitcoin, the block generating period is about 10 minutes, and the confirmation time can be 1 hour. Moreover, with proof of work, miners may compete to earn profits and dominate the blockchain with more CPU resources, which could lead to huge computation cost and energy consumption. Several alternative consensus mechanisms are proposed to prevent exhausting computation resources~\cite{kolb2020core,wang2018survey}.

Algorand is a recently proposed blockchain design that enables efficient block confirmation time. It is based on proof of stake, where users do not need to exhaust computation resources. Instead, users need to deposit money as the stake. The more money a user owns, the higher chance the user can be selected as a block proposer or verifier. Its security is guaranteed theoretically as long as the majority of stake is owned by honest users~\cite{chen2019algorand}. Furthermore, its scalability is demonstrated with simulations of up to 500,000 users, in which transaction confirmation time can be shortened to be within one minute \cite{gilad2017algorand}, making it feasible for use in some systems at real time. 
More specifically, for each round $r$, a leader $l^r$ and a small set of verifiers $SV^r$ that are randomly selected and publicized by cryptographic self-selection participate. The leader will build and propagate a new block, and then selected verifiers will reach the Byzantine agreement on it within a finite number of steps. Upon termination, the agreed block $B^r$ is added to the ledger. Since the block has already reached consensus and has been confirmed, the round period is exactly the transaction confirmation time. 

Our framework leverages Algorand, combining its protocol design with our ideas of designing dual cyber-physical blockchains at different time scales, using vehicles' physical sensing capabilities for verifying messages, and introducing the concept of proof-of-travel credits in stake computation.   
Moreover, to scale Algorand to CV applications in a large traffic network, we leverage the sharding technique~\cite{kolb2020core} for record transferring and sharing between blockchains in different regions. By choosing an appropriate region size, our framework design can have short round latency and low overhead, while preventing vehicles from transferring records frequently. 
In addition, our proposed summary step further reduces the overhead for new vehicles. 

\subsection{Blockchain in Transportation}
Blockchain has been leveraged by many researchers to advance applications in transportation systems in the literature.
\cite{pustivsek2016blockchain} presents a blockchain based negotiation process to select the most convenient electric vehicle (EV) charging station. 
\cite{gao2018blockchain} leverages blockchain to record EV charging activities, which enables information sharing while securing sensitive user information. 
\cite{yuan2016towards} presents a seven-layer conceptual model for blockchain in transportation and discusses ride sharing in the application layer.

There are also several works that consider using blockchain in vehicular networks.  
\cite{iqbal2019trust} reviews different models for calculating reputations of vehicles, discusses challenges of previously centralized and distributed methods, and indicates new directions on blockchain and fog computing. \cite{singh2018branch} proposes to secure vehicular communication by recording each message in blockchain, which is not scalable. Its proof of work scheme also leads to heavy computation burden.
\cite{yang2018blockchain} proposes a blockchain based trust management scheme for vehicular networks, in which road side units (RSUs) update and maintain the blockchain. Its consensus is partially based on proof of work, which still leads to high computation cost for every RSU, and its security in one region cannot be well protected once the RSU is compromised. 
\cite{dong2019blockchain} leverages Bitcoin-NG and hierarchical design to reduce transmission latency of blocks and increases the throughput capacity. However, this hierarchical design assumes that RSUs will maintain a higher-layer blockchain, which processes all information from lower-layer blockchains. It also has high computation resource demand and is not resilient to attacks. \cite{dorri2017towards} proposes a similar idea that may also suffer from the failure of one node. \cite{jiang2018blockchain} attempts to mitigate the resource demands by proposing concepts of sub-blockchains for different node types, e.g., vehicles, RSUs, and stations. However, it lacks implementation details such as communication across different sub-blockchains.

Our work addresses the challenge in applying blockchain to CV applications with a novel framework design that is introduced below in detail.

\section{Design of Our Framework}
\label{sec: methodology}

To avoid confusion, in our work, \emph{messages} carry traffic information that vehicles share with others, e.g., Basic Safety Message (BSM), Cooperative Awareness Message (CAM), Cooperative Perception Message (CPM), etc. 
To build trust among vehicles via blockchain, vehicles may generate \emph{transactions} and broadcast them to other vehicles in the same region. The transactions generally include the report of malicious misbehavior, application of transferring records, etc. By aggregating all transactions within a period, one vehicle can then update all vehicles' records and build a block accordingly.

\subsection{Framework Overview}
Our blockchain design leverages the block generation and consensus mechanism from Algorand. Technically speaking, it is based on proof of stake. In our design, the stake is computed from trust points and proof-of-travel credits. The higher stake a vehicle holds, the higher chance it can be selected as a leader or verifier to maintain the blockchains. When driving, only traffic information sent from vehicles with high trust points and proof-of-travel credits can be viewed as trustworthy. Such design sets up a relatively high threshold for malicious vehicles to manipulate the blockchains and start attacks. 

To update blockchain on time, and alleviate resource demand for vehicles, we leverage the sharding method to partition vehicles into subsets according to their traveling region. Each vehicle has a permanent address and a current active address. For trust points blockchain, attackers' misbehavior should be exposed as soon as possible, and surrounding vehicles in the current active region can respond within a short time. Thus, trust points blockchain in one region is maintained by those vehicles that claim to be active in the current region and will only record those vehicles' trust points. A vehicle that moves across different regions can claim its new active region and have its trust points record transferred. The detailed mechanism is introduced in Section~\ref{tp blockchain}. For proof-of-travel blockchain, since it records a vehicle's historical information over a long period (e.g., 100 days), it is maintained by those vehicles that have their permanent address in this region. It will only record those vehicles' proof-of-travel credits. In this way, different regions will have their blockchains maintained and updated independently. 


Other details, such as the transaction generation and physical verification processes for the two blockchains, are introduced in the following Sections~\ref{tp blockchain} and~\ref{pot blockchain}, respectively.


\subsection{Trust Points Blockchain}\label{tp blockchain}
Trust points blockchain is mainly for identifying and exposing malicious attackers. When one vehicle sends out a message, surrounding vehicles within the communication range can receive and verify it based on information from their own on-board sensors or other sources (e.g., a message with falsified position data may be deemed as suspicious by surrounding vehicles using their own sensors\footnote{How vehicles may verify messages based on physical information from their own sensors or other sources is beyond the scope of this paper, which focuses on the design of the blockchain framework.}). Then depending on the situation, various types of transactions can be generated and sent. Each transaction includes transaction ID (TID), smart contract type ID (SCID), senderID, debateID, regionID, location, time and payload. Transaction senders will encrypt the transactions with their private keys and broadcast them. Receivers can verify the identity of senders with the public keys. SCID identifies the transaction type and the corresponding smart contract for updating vehicles' points/credits. Location and time are the transaction generation location and time. DebateID is the ID of the vehicle that sends out the suspicious message. RegionID is the ID of the region that a vehicle is moving into. TID is the hash value of senderID, debateID, regionID, location and time, and is considered unique. 
Various transactions are sent in the following situations:

\begin{myitemize}
\item If a vehicle disagrees with a message sent by another vehicle, it can report this disagreement in a transaction with SCID=0000.
\item Upon receiving a disagreement transaction (SCID=0000), other vehicles can take a stand on agreeing or disagreeing in a new transaction with SCID=0000 if they have not done so.
\item If a vehicle disagrees with the judgement made in the previous voting process (details of the process are introduced later in Algorithm~\ref{contract:voting}), 
it can report the disagreement after enough evidence is gathered (i.e., when stake of honest vehicles gets larger) in a new transaction with SCID=0001.
\item A vehicle moving to another region may apply to transfer its records in a transactions with SCID=0002. 

\end{myitemize}


The transactions received within a period (i.e, the round latency of trust points blockchain) are processed with our designed contracts in Algorithms~\ref{contract:voting}, \ref{contract:redressing} and~\ref{contract:transfer}. 

\subsubsection{Instant voting contract}
Transactions with SCID=0000 are to report attackers and falsified messages. They are handled by an instant voting contract, as shown in Algorithm~\ref{contract:voting}. 

Let us use the highway merging application in Fig.~\ref{fig_attacks}(a) to help explain the algorithm.
First, the attacker near the ramp sends out a falsified message with wrong traffic status to mislead vehicles on the ramp. If there are honest vehicles in both the examination and communication ranges, they can also receive the message and may deem it suspicious based on the information from their own on-board sensors. They can then generate transactions with SCID=0000 to report such findings and broadcast the transactions to other vehicles in the region. Within a period, surrounding vehicles within both the communication and examination ranges are all supposed to take a stand. Then other vehicles in the region collect all transactions and start the smart contract based voting process as shown in Algorithm~\ref{contract:voting}. The voting process for this contract instance will classify all transaction senders into two groups. One group of vehicles agree with the message's content sent by the vehicle with debateID, and the other group of vehicles disagrees with that. The contract will make a final judgment and update vehicles' trust points by comparing the two groups' accumulated stake. 
The group with the higher stake will be called majority and have their trust points increased by 1, while the other group will be called minority and have trust points of -1.

\begin{algorithm}[h]
\SetAlgoLined
\KwResult{Updated trust points for involved vehicles}
 \textbf{Input: }\text{senderID, debateID, location, time, opinion}\\
 \If{time has not elapsed 2$tb_s$ since receiving the suspicious message ( $tb_s$ denotes the time bound that most vehicles in the same region can receive the transaction)}{
 Find an ongoing contract with the same debateID, different senderID, close location and time\;
 \If {no such contract}{
   Create a new contract instance for debateID\;
   }
   \eIf{opinion is agree}{
   Add senderID to the agree group\;
   }{
   Add senderID to the disagree group\;
   }}
   \If{any contract has existed 2$tb_s$}{
   Compare the stakes between the agree group and the disagree group\;
   Increase the trust points of the majority by 1\;
   Set the trust points of the minority to be -1\;
   }
 \caption{Instant Voting for SCID=0000 Transactions}
 \label{contract:voting}
\end{algorithm}

\subsubsection{Redressing contract}
A group of attackers with high stake may dominate the voting process if the number of surrounding honest vehicles is initially limited. To avoid repeated attacks from the group of attackers, we design the redressing contract, which allows re-evaluation of vehicles' opinions on the message content from the vehicle with debateID.  
In particular, if a vehicle disagrees with the previous voting result obtained by Algorithm~\ref{contract:voting}, it can send a transaction with SCID=0001, which triggers the redressing process in Algorithm~\ref{contract:redressing}. The redressing algorithm finds all the ended contracts that involve the debateID, and form groups $G_s$ and $G_o$ that represent all the vehicles agree and disagree with debateID, respectively. 
Let $N(G_s)$ and $N(G_o)$ denote the accumulated state for group $G_s$ and $G_o$, respectively. If the stake difference between the two groups is larger than a threshold $N_{th}$, 
previous judgement can be redressed. 

\begin{algorithm}[h]
\SetAlgoLined
\KwResult{Updated trust points for involved vehicles}
\textbf{Input: }\text{senderID, debateID, time}\\
 Find all ended contracts of type SCID=0000, in which the vehicle with debateID finally won\;
 From all those ended contracts, build a group $G_s$ that includes all vehicles that agree with debatedID and $G_o$ that includes all vehicles disagree with debated ID\;
Let  $N(G)$ denote the accumulated stake for vehicles in group $G$ and $N_{th}$ denote the threshold number\;
  \If{$N(G_o)-N(G_s)>N_{th}$}{
   Redress previous contract instances\;
   Increase trust points for group $G_o$ by 1\;
   Set trust points for group $G_s$ to be -1\;
   }
 
 \caption{Redressing for SCID=0001 Transactions}
 \label{contract:redressing}
\end{algorithm}

\subsubsection{Transferring records contract}
When a vehicle moves across different regions, it is required to update its current active region. This way the vehicles within the same active region can communicate efficiently and the real-time performance can be improved. 
To update the active region, a vehicle needs to transfer trust points and copy its proof-of-travel credits to the new region. This process is shown in Algorithm~\ref{contract:transfer}. Note that the trust points of this vehicle in the original region should be set to zero.

\begin{algorithm}[h]
\SetAlgoLined
\KwResult{Updated trust points for senderID in two regions}
\textbf{Input: }\text{senderID, regionID, position}\\
  \If{position is near the border of current region and the region of regionID}{
   Get proof-of-travel credits and trust points of senderID in current active region\;
 Transfer these records to another region corresponding to regionID\;
 \If{another region updated records of senderID}{Set trust points of the vehicle to 0 in original region}
   }
 \caption{Transferring Records for SCID=0002 Transactions}
 \label{contract:transfer}
\end{algorithm}

\subsection{Proof-of-Travel Blockchain}\label{pot blockchain}
Proof-of-travel blockchain is mainly to record vehicles' accumulated contributions to other vehicles. 
We assume that CVs will broadcast traffic information, e.g., BSM or CAM messages, in an average frequency of $f_m$ Hz. Surrounding vehicles within the communication range will record the public keys of message senders and the number of valid messages. After each travel period $T_{pot}$ (which is much longer than the period for trust points blockchain), every vehicle $v_i$ will have a list that records the public keys of message senders $v_j$ and the number of messages $n_{j2i}$ from every sender during this period. We propose a metric $n^{pot}_j$, named ``proof-of-travel credits'', to summarize all messages commenting on vehicle $v_j$ from all other vehicles $v_i$, i.e., 

\begin{equation}\label{proof_of_travel}
n^{pot}_j=\sum_{i}~n_{j2i}
\end{equation}

For a vehicle, its accumulated proof-of-travel credits in the latest $N_{sum}$ periods is computed by:
\begin{equation}\label{sum_proof_of_travel}
N^{pot}_j[m]=\sum^{N_{sum}-1}_{k=0}\alpha^{k}~n^{pot}_j[m-k]
\end{equation}
where $n^{pot}_j[m-k]$ is the proof-of-travel credits in the $(m-k)_{th}$ period, $\alpha$ is a discounting factor, $N^{pot}_j[m]$ is the accumulated $N_{sum}$ periods proof-of-travel credits by the $m_{th}$ period. Vehicles that make a persistent contribution to others will have a high accumulated proof-of-travel credits. 

We assume that most vehicles will report the number of messages they received honestly because honest vehicles generally do not know the identity of surrounding vehicles. 
Honest vehicles can also generate transactions to reveal falsified number report, in a similar way as in the trust points blockchain. 

\begin{figure}[t]
\centering\includegraphics[scale=0.25]{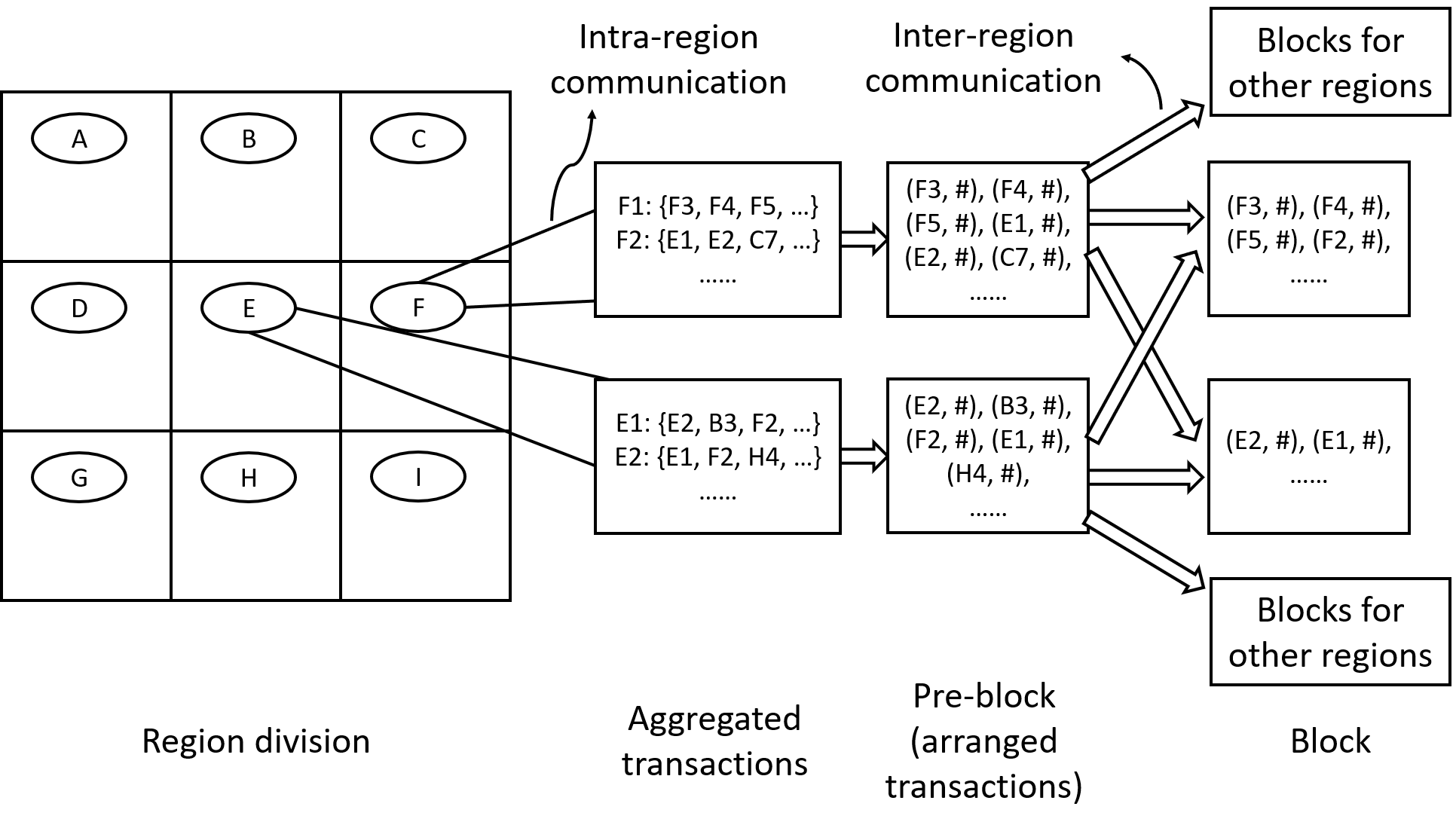}
\caption{Proof-of-travel blockchain updating across multiple regions. \textmd{With intra-region communication, each vehicle can aggregate transactions received from others in the same region. By computation and arrangement, vehicles can build a pre-block and reach consensus on it, which includes vehicle ID and proof-of-travel credits. Note that records in pre-blocks may not be complete as vehicles may move across multiple regions; but with inter-region communication, blocks for each region will be eventually built.}}
\label{fig: pot}
\end{figure}

Fig.~\ref{fig: pot} presents the general process of updating the proof-of-travel blockchain. Different vehicles may get registered in different permanent regions, e.g., region A, region B, and so on. We assume that each vehicle is uniquely identified by the regionID and the vehicle index in that region, e.g., A1 is a vehicle with the index of 1 in region A. Since a vehicle may change lane, overtake, choose different routes, or even enter another region, it will encounter different vehicles in the same or different regions. It will record those vehicles and the number of messages they sent in a transaction and broadcast it in the region. As in the figure, each region will have a number of aggregated transactions. Each transaction is denoted by the vehicle senderID with a set including all vehicles it encountered, e.g., E1: \{E2, B3, F2, $\cdots$\} is the transaction sent by vehicle E1. The selected block proposer will then arrange these transactions and build a pre-block by summarizing credits for all vehicles in the sets. After the consensus process, all vehicles in the region will agree with this pre-block. We name it pre-block because it may not include complete information for vehicles in the region, but record vehicles who are not in this region. Then vehicles in neighboring regions will communicate with each other regarding the pre-blocks. Finally block proposers will build a block containing complete information of vehicles in their region. In this way, to find the proof-of-travel credits of one vehicle, we can easily refer to the blockchain in that vehicle's permanent region.

\subsection{Stake Computation and Region Size}\label{stake_computation}

In framework, the stake $N^{stk}$ is computed from the accumulated proof-of-travel credits $N^{pot}$ in the proof-of-travel blockchain and the trust points $N^{tp}$ in the trust points blockchain:

\begin{equation}\label{stake}
N^{stk}=\frac{N^{pot}}{1+M^{pot}}+\frac{N^{tp}}{1+M^{tp}}
\end{equation}
where $M^{pot}$ and $M^{tp}$ denote the mean value of $N^{pot}$ and $N^{tp}$ for all vehicles, respectively. Here we use $1+M^{pot}$ and $1+M^{tp}$ as denominator to avoid the near-zero case. 
This stake is the basis of consensus mechanism for both the trust points blockchain and the proof-of-travel blockchain.

The region size that one blockchain can cover is decided by various factors. First, it is limited by the round latency that we desire. A larger size may result in a larger latency to reach consensus, thus slow down system's response to attacks. Second, for trust points blockchain, the smaller the region is, the lower the resource demand is for all vehicles in the region to maintain and update the blockchain. However, a small region size typically means that vehicles will frequently move across different regions, thus frequently transferring their records. For proof-of-travel blockchain, a smaller region size leads to lower overhead on intra-region communication but higher overhead on inter-region communication. Such trade-offs are addressed in detail in Section~\ref{subsec: round latency} and~\ref{sec: resource demand}.

\section{Security Analysis}\label{sec: defense}


As discussed in the threat models in Section~\ref{subsec:threat_model}, a malicious vehicle can broadcast falsified messages, report falsified misbehavior of honest vehicles, or 
create other pseudonymous identities for manipulating the voting process.

We evaluate the effectiveness of our proposed framework in defending against the three threat models, based on simulations in SUMO. We also analyze the relationship between round latency of the trust points blockchain and the region size in our framework.



\subsection{Against Message Spoofing Attack}

For the threat model described in Fig.~\ref{fig_attacks}(a), under message spoofing attack and without our framework, the merging vehicle in on-ramp can be fooled to speed up when the highway is congested or to decelerate when there are few vehicles on the highway. When the merging vehicle is able to perceive the traffic environment, it has to adjust its velocity significantly, which leads to low efficiency and possible failure to merge onto the highway. With the protection of our framework, other honest vehicles on the highway may report this spoofing attack and launch the voting process to assess the message. Before the assessment is finished, the merging vehicle can start taking cautious actions in milliseconds, e.g., speeding up only a little bit for reacting to uncertain traffic status on the highway, preventing sudden velocity changes. When all vehicles reach consensus after the voting process, the identity of the attacker is exposed, and its messages are not trustworthy any more.



We compare the average speed, CO emission, and fuel consumption of the merging vehicles under three cases, i.e., without attack, under attack without our framework, and under attack with our framework. We perform 20 simulations for each case, and the average performance is recorded in Table~\ref{table:spoofing}, where `ours' is short for our framework. During simulations, vehicle arriving rate on highway and on-ramp are both 0.2 vehicle per second. The performance is measured over the 200 meters on-ramp and the following 200 meters highway. 
The table shows that without the protection from our framework, the attack can lead to a 42.66\% decrease in average speed and an 87.06\% increase in CO emission. With our framework, the vehicles can take cautious actions, thus have a significantly higher travel speed and lower CO emission.

\begin{table}[h]
\centering
\caption{Effectiveness of our framework in protecting against message spoofing attack in highway merging.}
\label{table:spoofing}
\begin{tabular}{c|C{1cm}|C{1.8cm}|C{1.6cm}}
\hline
performance             & without attack & under attack without ours  & under attack with ours \\ \hline
average speed ($m/s$)   & 16.41          & 9.41                      & 13.00                  \\
CO ($mg$)               & 741.06         & 1386.24                   & 1006.36                \\
fuel ($mL$) & 41.17          & 41.91                     & 42.96                  \\ \hline
\end{tabular}
\end{table}




\subsection{Against Bad Mouthing Attack}
\begin{figure}[t]
\centering\includegraphics[scale=0.34]{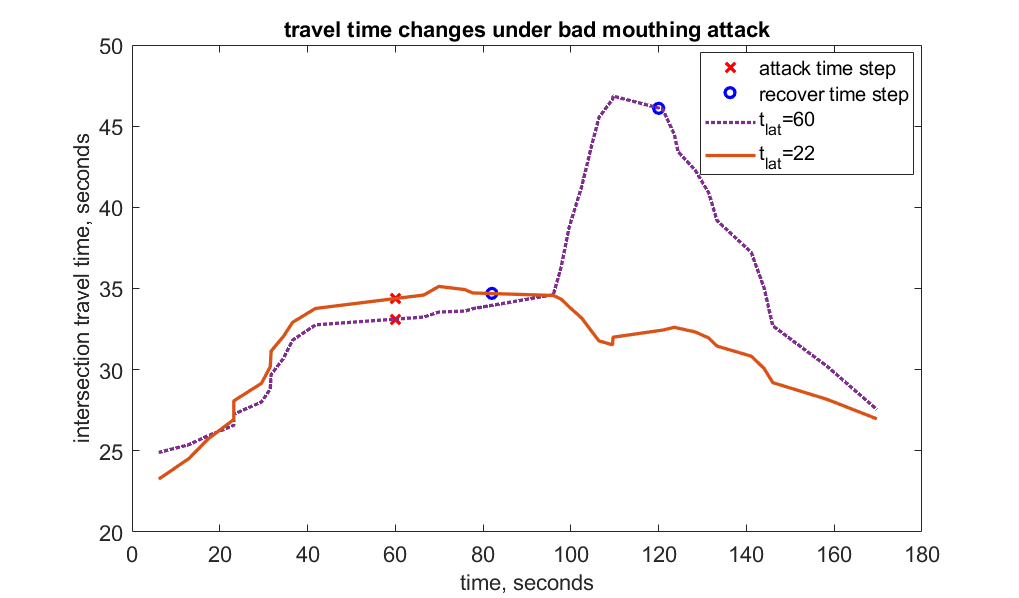}
\caption{Effectiveness of our framework against bad mouthing attack in an intelligent intersection. \textmd{The red stars denote the moment that the attack starts and the blue circles denote the moment that our framework detects the attack and the traffic system starts recovering from the attack. Travel time curves with different round latency ($t_{lat}=22 \text{ and } 60$ seconds) of the trust points blockchain are plotted.
}}
\label{fig:bad_mouthing}
\end{figure}

Under the bad mouthing attack as shown in Fig.~\ref{fig_attacks}(b), all honest vehicles near the intersection are reported to have misbehavior by the attacker. Those honest vehicles may generate transactions to claim that they are innocent and being attacked. Before it is confirmed, all vehicles near the intersection will take cautious actions and choose not to trust others' messages; thus, the intelligent intersection may temporarily lose its functions. Vehicles will travel through the intersection as if there were all-way stop signs. 

Fig.~\ref{fig:bad_mouthing} shows the traffic in an intelligent intersection that is under attack. The vehicle arriving rate is 0.05 vehicle per second in the intersection. Around time 60 (seconds), the attacker starts the bad mouthing attack. However, the design of our framework ensures that the attack can be soon discovered and the system can return to normal shortly. Specifically, in one round of trust points blockchain, the surrounding vehicles have reached consensus on the existence of this attack. An honest vehicle will then trust messages from other honest vehicles, and the function of intelligent intersection gets recovered. From the figure, we can see that if our framework has a smaller round latency of $t_{lat}=22$ seconds for the trust points blockchain, the system can recover before any significant increase of travel time. If the round latency reaches $t_{lat}=60$ seconds, travel time starts increasing initially when the intelligent function is temporarily disabled, but starts decreasing around time 120 and fully recovers around time 140. More analysis of the impact of round latency is shown later in Section~\ref{subsec: round latency}.

\subsection{Against Sybil and Voting Attack}
\begin{figure}[t]
\centering\includegraphics[scale=0.34]{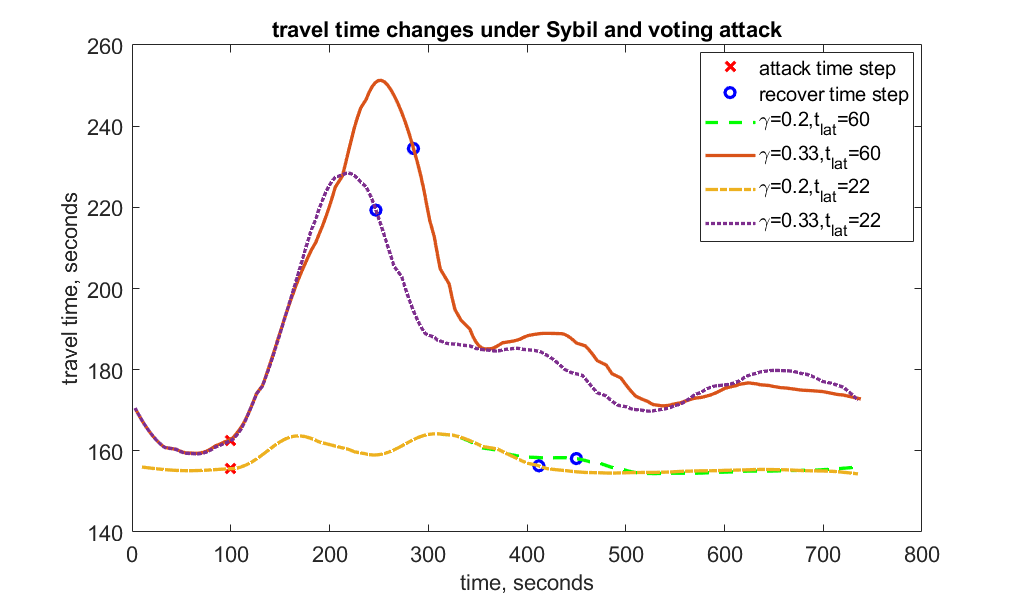}
\caption{Effectiveness of our framework against Sybil and voting attack. \textmd{Red stars denote the moment the attack starts, and the blue circles denote the moment the attack is detected and the recovery starts. Travel time curves with different round latency ($t_{lat}=22 \text{ and } 60$ seconds) for the trust points blockchain and under different vehicle arriving rates ($\gamma=0.2 \text{ and } 0.33$ vehicle per second) are plotted.}}
\label{fig:sybil}
\end{figure}


Fig.~\ref{fig_attacks}(c) shows the threat model where a traffic network with route choices is attacked with Sybil and voting attack. Specifically, an attack and the two Sybil vehicles it generates send transactions reporting a traffic accident in one route and win the voting process because they have more stake initially. 
Then other vehicles arriving at the area will have a higher probability (e.g., 0.9 in our simulations) to take another route, which leads to longer travel time. However, as several vehicles still choose the route with the fake accident and realize that it is a Sybil and voting attack, they trigger the redressing process as the honest vehicles have higher stake now. 
Fig.~\ref{fig:sybil} shows that our framework can soon discover the Sybil and voting attack, and recover to normal status. We also analyze the impact of vehicle arriving rate and round latency of trust points blockchain, and observe that a higher vehicle arriving rate leads to longer travel time under attack but faster recovery time in our framework. A smaller round latency may reduce the travel time under attack and also shorten the recovery time.

\smallskip
\noindent
The results above demonstrate the effectiveness of our framework in protecting against message spoofing, bad mouthing, and Sybil and voting attacks. 
Note that with our proof-of-travel blockchain, it is even harder to perform these attacks (especially Sybil attack), as it takes more effort for the malicious attackers to build up a travel record in the region with high stake.





\subsection{Round Latency and Region Size}\label{subsec: round latency}

In the results above, we have seen the impact of round latency of the trust points blockchain on system performance. This latency is closely related to the region size. Here we conduct more in-depth analysis on the relationship between the two.



In~\cite{chen2019algorand} and~\cite{gilad2017algorand}, the authors simulated 50k Algorand users in the m4.2xlarge virtual machines on Amazon's EC2 platform, with 50 users per VM. The message transmission in Algorand is enabled by gossip protocol, which is similar to Bitcoin. The bandwidth for each Algorand process is set to 20 Mbps. Latency for one round of Algorand is about 22 seconds, dominated by the time to gossip a 1 MB block through the user network~\cite{gilad2017algorand}.

We assume that the blockchain will record all transactions sent by vehicles in a circular region with a radius of $d_0$. The size of a block is computed as:
\begin{equation}\label{block_size}
S_{b}=(\frac{2\pi d_0 \beta_l \beta_c}{\beta_v} + \pi d_0^2 \beta_d \beta_t)t_{lat} S_t^{SCID}
\end{equation}
where $\beta_l$ is the number of lanes connect to outer space per perimeter of the region, $\beta_c$ is the traffic capacity of a single lane with the unit as the number of vehicles traveled per hour per lane, $\beta_v$ is the number of vehicles that can be claimed in a transaction of type SCID=0002, $\beta_d$ is the density of vehicles in this region, $\beta_t$ is the generation rate of transactions of type SCID=0000/0001, $t_{lat}$ is the latency (period) to generate a block, $S_t^{SCID}$ is the size of a transaction.

We set the capacity of lanes that connect to other regions, $\beta_c$, to be 3000 vehicles per hour per lane, as observed in the data collected from a section of the highway I-5 in Southern California~\cite{caltrans} and should be similar in other regions. 
The density of vehicles $\beta_d$ is set to 300 veh/$\text{km}^2$, which is similar to the density in Beijing~\cite{yang2014review}. Considering that attacking events are rare and vehicles are not driving on the road all the time, $\beta_t$ is set to 0.05 transactions per hour per vehicle. $S_t^{SCID}$ is 250 byte per transaction. $\beta_v$ is set to 10 vehicles per transaction and $\beta_l$ is 1 lane per km.





\begin{figure}[t]
\centering\includegraphics[scale=0.32]{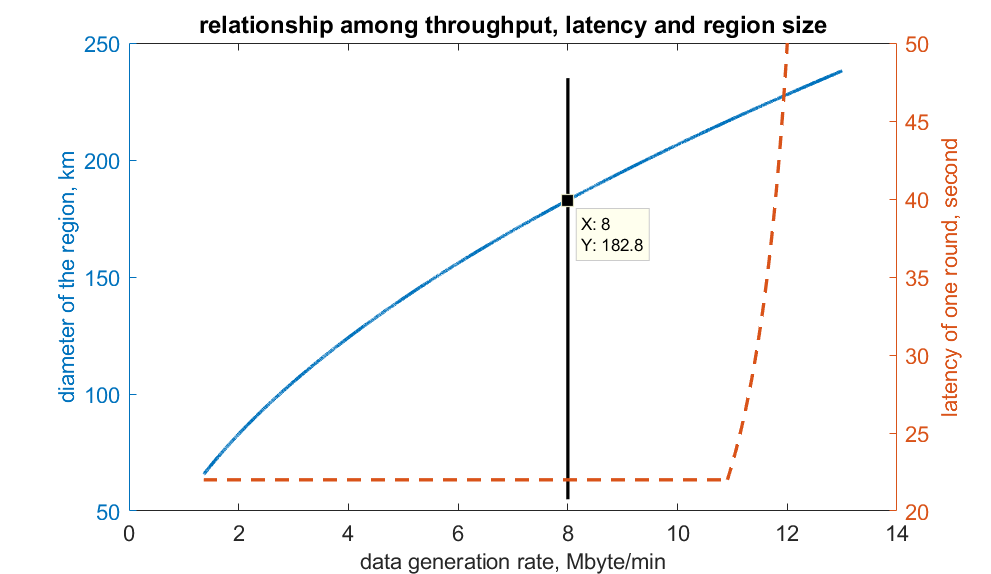}
\caption{Maximum region size and minimum round latency under different transaction generation rate (best viewed in color). \textmd{The solid blue line and the dashed orange line represent the maximum region size and the minimum round latency under different transaction generation rate, respectively. Drawing a vertical line, we can see the minimum round latency corresponding to the region size, e.g., 22 seconds for a region with a radius of 182.8 km.}}
\label{fig_commu}
\end{figure}

Taking a region with a similar size to Beijing, the transaction data generation rate is about 1.6 Mbyte/min. 
According to~\cite{gilad2017algorand}, the latency for one round of Algorand has little variation on the number of users. It basically remains the same when the block size is smaller than 4MB and then increases proportionally with a larger block size. We plot the relationship among throughput (data generation rate), minimum latency, and region size in Fig.~\ref{fig_commu}, and we can clearly see the trade-offs between the minimum round latency and the maximum region size. 

\section{Resource Demand Analysis}
\label{sec: resource demand}


We use three metrics to measure the resource demand of our framework on computation, communication, and storage cost. Here, the computation cost is the CPU resource used for running blockchain on each vehicle. Communication cost is measured by the total size of messages received per minute by each vehicle. 
Storage cost is measured by the total size of data maintained in our framework. 

We leverage the results from Algorand in our analysis. 
The Algorand users use CPU resources mainly for verifying signatures and verifiable random functions (VRFs). Each Algorand process uses about 6.5\% of a core averagely, in a 2.4 GHz Intel Xeon E5-2676 v3 (Haswell) Processor. While the on-board computing platform for vehicles may be different from the ones used in Algorand's analysis, we hope the overall trends shown in this section could still provide some valuable observations.


Next, we will analyze the three metrics for the trust points blockchain in Section~\ref{subsec:comp}, \ref{commu_cost}, and \ref{subsec:stor}, and briefly discuss the resource demand for the proof-of-travel blockchain in Section~\ref{subsec:POT}.

\subsection{Computation Cost}\label{subsec:comp}

\begin{figure}[t]
\centering\includegraphics[scale=0.35]{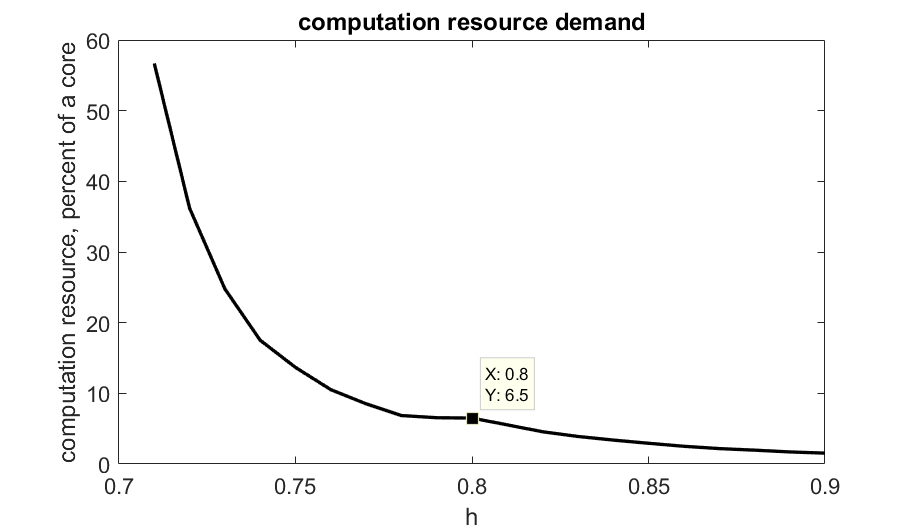}
\caption{Computation resource demand under different fractions of honest vehicles.}
\label{fig_compu}
\end{figure}

Note that the computation cost mainly depends on the verifier committee size, rather than the total amount of vehicles, because only block proposers and verifiers will broadcast messages. 
According to~\cite{gilad2017algorand}, the verifier committee size is directly computed from the fraction of honest vehicles $h$ and a parameter $F$. Here we set $F=5\times10e-9$, denoting a negligible probability that the verifier committee reaches consensus on a falsified block~\cite{gilad2017algorand}. By assuming that the computation cost is proportional to the number of messages received, we plot Fig.~\ref{fig_compu}, which shows the relationship between computation cost and the fraction of honest vehicles $h$. Here the computation cost is measured by the utility of one core in a 2.4 GHz Intel Xeon E5-2676 v3 (Haswell) Processor, e.g., it is 6.5 percent when 80 percent of the vehicles are honest.


\subsection{Communication Cost}\label{commu_cost}
According to the consensus process described in \cite{chen2019algorand} and \cite{gilad2017algorand}, communication cost in one round is computed as
\begin{equation}\label{commu}
C_{cm}=S_{b}+S_{c}\eta_{p}+S_{c}\eta_{v}(\eta_{s}-2)
\end{equation}
where $S_{b}$ and $S_{c}$ are the size of a block and a short message for every step in the consensus process, respectively. As in \cite{gilad2017algorand}, $S_c$ is assumed to be no more than 200 bytes. Let $\eta_{p}$ denote the number of block proposers in every round. All these proposers will broadcast a short message to claim their identity in Algorand. Let $\eta_{v}$ and $\eta_{s}$ denote the number of verifiers and the number of steps to reach consensus. Then $\eta_{v}(\eta_{s}-2)$ is the number of messages sent by verifiers in one round because they do not send messages in the first and last steps. 



\begin{figure}[t]
\centering\includegraphics[scale=0.32]{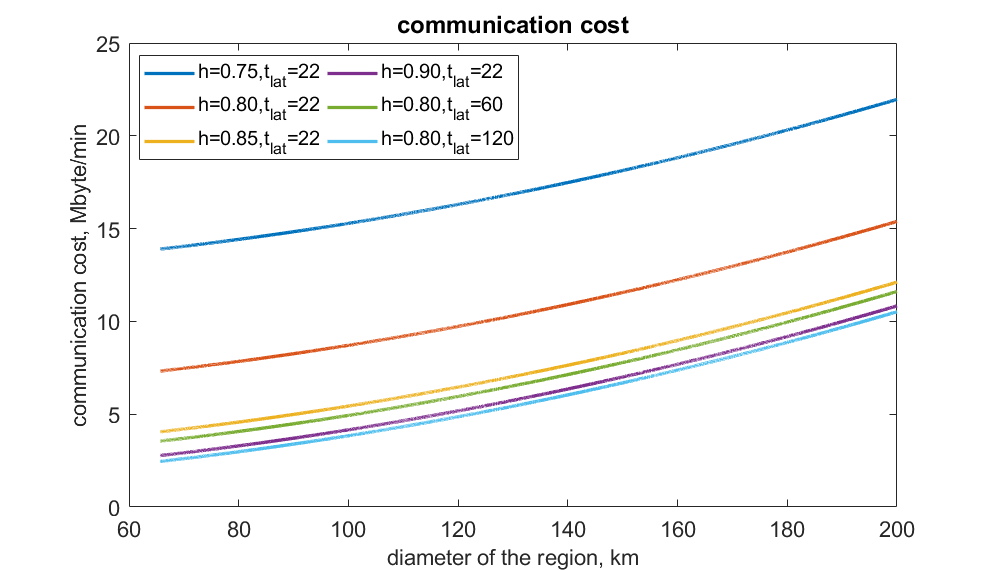}
\caption{Communication cost under different fraction of honest vehicles ($h$), round latency ($t_{lat}$), and region size.}
\label{fig_commu2}
\end{figure}

We plot communication cost under different fraction of honest vehicles, different round latency, and different region size in Fig.~\ref{fig_commu2}. We can see that the communication cost will increase with a smaller fraction of honest vehicles, a smaller round latency, or a larger region size.


\subsection{Storage Cost}\label{subsec:stor}

Storage resource needed by every vehicle for one-day data can be computed as
\begin{equation}\label{stor_1d}
C_{s}^{1d}=\frac{24\times60}{t_{lat}\alpha_s}S_{b}
\end{equation}
where $\frac{24\times60}{t_{lat}}$ is the number of blocks generated every day, and $\alpha_s$ is a sharding parameter. For N shards, vehicles only store blocks whose round number equals their public key modulo N, so that storage cost can be reduced by the divisor of $\alpha_s$.


As we mentioned in Section~\ref{subsec: round latency}, for a region with the similar size to Beijing, the blockchain generates 1.6 Mbyte data per minute, which means 69.12 Gbyte per month. Even if we leverage the sharding technique with $\alpha_s=10$, blockchain users will still have to maintain about 7 Gbyte of new data every month. 
Moreover, the sharding technique will not help new vehicles. A new vehicle will have to download all the data to get the correct status of all vehicles. 

To overcome this problem, we propose a new summary step for blockchain. In our design, selected block proposers will summarize the status of all vehicles in new blocks with a period of $t_{sum}$. By indicating the type of block in its header, a new vehicle can easily find the latest summary blocks and do not need to download blocks generated before those.

\begin{figure}[t]
\centering\includegraphics[scale=0.32]{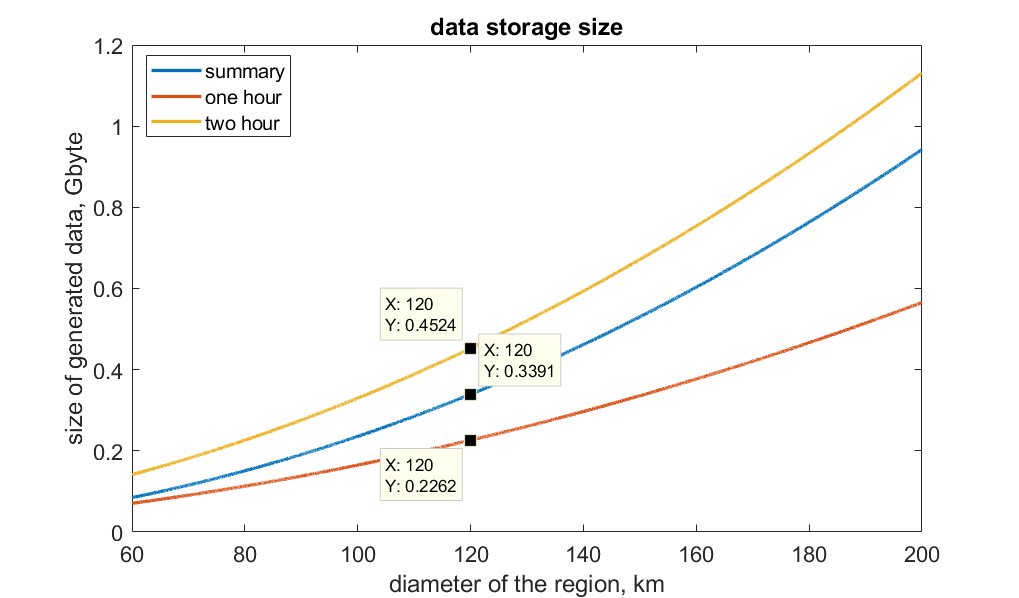}
\caption{Storage cost under different region size. \textmd{We can see that the summary of all vehicles' records has a size that is between the sizes of transactions generated in one hour and in two hours.}}
\label{fig_storage}
\end{figure}

In this way, without considering the sharding technique, storage resource demand $C_s$ is
\begin{equation}\label{stor_dem}
C_s=S_{b}+t_{sum}C_{s}^{1d}
\end{equation}
where $S_b$ is the data size of summary blocks, as computed below:
\begin{equation}\label{stor_sum}
S_{b}=\pi d_0^2 \beta_d S_u
\end{equation}
where $S_u$ denotes the size of data for summarizing one vehicle's status. 
As shown in Fig.~\ref{fig_storage}, the size of these summary blocks is between the size of transactions generated in one hour and the size of transactions generated in two hours. 

These summary blocks are communication overhead. A smaller $t_{sum}$ means larger communication overhead, smaller storage cost, and smaller size of data a new vehicle needs to download to join the blockchain. Taking a region with a radius of 120 km for example and reading data from Fig.~\ref{fig_storage}, with a period of $t_{sum}=24$ hours, the communication overhead would be about $\frac{0.339}{0.2263\times24}=6.24\%$, storage cost would only be $24\times0.2263+0.339=5.77$ Gbyte. That is, a new vehicle may only need to download 5.77 Gbyte data to get the latest status of blockchain.

\subsection{Resource Demand for Proof-of-Travel Blockchain}\label{subsec:POT}

To alleviate resource demand for the proof-of-travel blockchain, vehicles can generate blocks and update the blockchain in a much larger period than that of the trust points blockchain, e.g., one day or longer. Considering that vehicles usually stay most time in their permanent address/region, proof-of-travel blockchain is maintained by vehicles in the same permanent region to mitigate communication overhead.

For example, for a region with a 10 km radius, there are about 100 thousand registered vehicles under density $\beta_d$. We assume that a vehicle can record at most 500 vehicles with higher credits in its transaction, and the updating period is one day. The size of one transaction is estimated as $20\times500=10k$ bytes if each vehicle record consumes 20 bytes. The size of aggregated transactions will be $100k\times10k=1G$ bytes. After arrangement and inter-region communication, the final block should have a much smaller size. In this way, with a block generating period of one day or longer, the resource demand for the proof-of-travel blockchain should be much less than that for the  trust points blockchain.

\section{Conclusion}\label{sec: conclusion}
In this work, we propose a dual cyber-physical blockchain framework that includes a trust points blockchain and a proof-of-travel blockchain for building trust and securing communication for CV applications. The trust points blockchain has a quick response to suspicious behavior, with smart contracts designed for instant voting, redressing, and transferring records. 
The proof-of-travel blockchain builds up reputations from vehicles' long-term travel records. The trust points and the proof-of-travel credits are used together to compute the vehicle stake for running the consensus mechanisms in both blockchains.
Experimental results demonstrate the effectiveness of our framework in protection against message spoofing, bad mouthing, and Sybil and voting attacks, in representative CV applications. We also conducted preliminary analysis on the framework's resource demand. As this is the first step towards building a trust framework for CV applications, we plan to continue improving the efficiency of our framework and its prototyping. 
\bibliographystyle{IEEEtran}
\bibliography{main}

\end{document}